\documentclass{aa}
\usepackage{graphicx}
\usepackage[]{natbib}
\usepackage[]{lscape}
\bibpunct{(}{)}{;}{a}{}{,}
\newcommand{\etal}{{\it et al.}}
\newcommand{\Msun}{\hbox{M$_\odot$}}

\newcommand{\kms}{\mbox{km~s$^{-1}$}}
\newcommand{\mhi}{\mbox{M$_{HI}$}}
\newcommand{\mmol}{\mbox{M$_{mol}$}}
\newcommand{\md}{\mbox{M$_{dust}$}}

\newcommand{\area}{\mbox{D$^2_{25}$}}
\newcommand{\LB}{\mbox{L$_B$}}
\newcommand{\LFIR}{\mbox{L$_{FIR}$}}
\newcommand{\Lx}{\mbox{L$_X$}}

\def\sec{\ifmmode{^{\prime\prime}}\else{$^{\prime\prime}$}\fi}
\def\min{\ifmmode{^{\prime}}\else{$^{\prime}$}\fi}
\def\deg{\ifmmode{^\circ}\else{$^\circ$}\fi}
\def\arcsec#1.#2 {\ifmmode {#1^{\prime\prime}\hskip-0.42em.
                  \hskip0.15em#2}
         \else {$#1^{\prime\prime}\hskip-0.42em.\hskip0.15em#2$}
         \fi}
\def\arcmin#1.#2 {\ifmmode {#1^{\hskip 0.05em\prime}\hskip-0.35em.
                  \hskip0.05em#2}
         \else {$#1^{\hskip 0.05em\prime}\hskip-0.35em.\hskip0.05em#2$}
         \fi}
\def\arcdeg#1.#2 {\ifmmode {#1\deg\hskip-0.42em.
                  \hskip0.10em#2}
         \else {$#1\deg\hskip-0.42em.\hskip0.10em#2$}
         \fi}

\begin{document}

\title{The gas content of peculiar galaxies: strongly interacting systems\thanks{Table 1 
is only available in electronic form at the CDS via anonymous ftp to cdsarc.u-strasbg.fr 
(130.79.128.5) or via http://cdsweb.u-strasbg.fr/cgi-bin/qcat?J/A+A/}}

\author{V. Casasola\inst{1}
           \and D. Bettoni \inst{2}
           \and G. Galletta\inst{1}
                         }

\offprints{G. Galletta}

\institute{Dipartimento di Astronomia, Universit\`a di Padova, Vicolo
               dell'Osservatorio 2, I-35122, Padova\\
             \email{galletta@pd.astro.it; casasola@pd.astro.it}
         \and Osservatorio Astronomico di Padova, Vicolo dell'Osservatorio 5,
        I-35122, Padova\\
             \email{bettoni@pd.astro.it}
             }

\date{Received ; accepted  }

\titlerunning{Interacting Galaxies}

\abstract{A study of the gas content in 1038 interacting galaxies, essentially selected from Arp, 
Arp and Madore, Vorontsov-Velyaminov catalogues and some of the published literature, is presented here. 
The data on the interstellar medium have been extracted from a number of sources in the literature 
and compared with a sample of 1916 normal galaxies. The mean values for each of the different ISM tracers 
(FIR, 21 cm, CO lines, X-ray) have been estimated by means of survival analysis techniques, 
in order to take into account the presence of upper limits. 
From the data it appears that interacting galaxies have a higher gas content than normal ones.
Galaxies classified as ellipticals have both a dust and gas content one order of magnitude higher
than normal. Spirals have in most part a normal dust and HI content but an higher molecular 
gas mass. The X-ray luminosity also appears higher than that of normal galaxies of same morphological type,
both including or excluding AGNs. 
We considered the alternative possibilities that the molecular gas excess may derive from
the existence of tidal torques which produce gas infall from the surrounding regions or from
a different metallicity which affects the X conversion factor between the observed CO line
luminosity and the H$_2$ calculated mass. According to our tests, it appears that interacting
galaxies possess a higher molecular mass than normal galaxies but with a similar star formation 
efficiency.
}

\maketitle

\section{Introduction}
It is known that interactions between galaxies and environment play an 
important role in determining the internal galaxy structure. Gas accretion 
may produces central black holes \citep{bh}, polar rings \citep{pr} or counter-rotations 
\citep{crgg, crbis}. Gas stripping in clusters may deplete a spiral galaxy transforming it 
into an S0 \citep{dressler97}, while massive collisions may destroy the whole galaxy
structure creating giant triaxial ellipticals \citep{bendo}. According to numerical models,
bars and rings may also be generated by a close encounter between galaxies \citep{lia}. 
Interaction also affects other galaxy properties, such as the star formation rate 
\citep{sanders,sage88,combes}, visible as a strong increase of infrared luminosity.

Inside galaxies, gas is expected to reflect the effects of the interaction more strongly 
than stars. A recent study on the interstellar medium (ISM) in 104 peculiar 
galaxies \citep{polar} shows that polar ring galaxies have a gas content one order of 
magnitude higher than normal ones. In many cases, the higher gas content is 
visible even if the interaction with the environment has ceased for long time.

We wonder if a peculiar gas content is detectable in those galaxies where interaction is 
currently ongoing, such as some systems contained in the \citet{vv}, \citet{arp} and
\citet{am} catalogues. Here we present the results of a selection of 1038 objects 
extracted from these catalogues. For these galaxies data on dust, HI, molecular  gas and 
X-ray luminosities are available in the literature. The properties of the different components 
of the interstellar medium are studied here and compared with those of a sample of 1916
normal galaxies 
\citep{normal}.
 
\setcounter{table}{1}
\begin{table}                                              
\caption{Number of interacting galaxies observed (N) and detected (N$_d$) according to the different ISM 
tracers. The column labeled `All' represents the number of galaxies present in each morphological type 
bin. The type 11 is assigned here to galaxies whose structure is not attributed to a specific morphological type in 
LEDA. }                                              
\tabcolsep 0.15truecm
\begin{center}
\begin{tabular}{lrrrrrr}                                              
\hline                                              
\hline                                              
	&		& All	 & \multicolumn{1}{c}{Dust} &	\multicolumn{1}{c}{HI}	& \multicolumn{1}{c}{Mol} &	\multicolumn{1}{c}{X-ray}	\\
Type	&	t	& 	 &  N/N$_d$	& N/N$_d$	& N/N$_d$	& N/N$_d$	\\
\hline                                              
E 	&	-5 	&  22  & 15/8 	& 	12/7 	& 	3/0 	& 	13/11	\\ 
E$^+$ & 	-4 	&  21  & 18/12 	& 	6/3 	& 	3/1 	&	11/10	\\ 
E/S0 & 	-3 	&  18  & 13/9 	& 	9/5 	& 	4/1 	&	9/4 	\\ 
S0 & 		-2 	&  47  & 36/31 	& 	21/14	& 	8/2 	& 	11/7 	\\ 
S0$^+$ &	-1 	&  40  & 35/31 	& 	18/15	& 	3/2 	& 	4/1 	\\ 
S0/a & 	 0 	&  36  & 31/25 	& 	14/14	& 	7/4 	& 	3/1 	\\ 
Sa 	& 	 1 	&  82  & 75/71 	& 	44/41	& 	12/12	& 	9/5 	\\ 
Sab 	&	 2 	&  78  & 67/63 	& 	42/42	& 	10/9 	& 	6/3 	\\ 
Sb 	& 	 3  	& 142  & 128/127 	& 	75/72	& 	29/23	& 	8/5 	\\ 
Sbc 	& 	4  	& 177  & 169/164 	& 	68/64	& 	16/15	& 	10/8 	\\ 
Sc+S? & 	5 	& 148  & 129/124 	& 	89/87	& 	19/17	& 	10/3 	\\ 
Sc 	& 	6  	&  74  & 63/58 	& 	50/48	& 	12/11	& 	8/2 	\\ 
Scd 	& 	7 	&  33  & 31/30 	& 	20/20	& 	4/4 	& 	5/4 	\\ 
Sd 	& 	8 	&  30  & 23/20 	& 	26/26	& 	4/2 	& 	2/1 	\\ 
Sm 	& 	9 	&  30  & 22/19 	& 	25/24	& 	10/6 	& 	4/3 	\\ 
Irr 	& 	10 	&  35  & 23/23 	& 	30/29	& 	5/5 	& 	4/3 	\\ 
? 	& 	11 	&  25  & 22/21 	& 	7/7 	& 	4/4 	& 	1/0    \\ 
\hline
Total  &  		& 1038 & 900/836 	& 556/518 	& 153/118  	& 	118/71 \\ 
\hline
\hline
\end{tabular}                                              
\end{center}
\label{numbers}                                              
\end{table}                                              

\section{Data extraction}
Our first effort was to define a list of galaxies appearing to be clearly interacting with 
nearby objects and presenting tidal tails or bridges. Merging systems and galaxies with 
disturbed structures have also been included in our sample. The first catalogues of 
galaxies of this kind were compiled by \citet{vv}, \citet{arp} and \citet{am}. 
These three catalogues represent almost the totality of our sample galaxies. 
However, all the above catalogues also contain objects that have been considered 
peculiar by the authors because of an unusually low surface brightness or 
the presence in a elongated group or chain of galaxies. The morphology 
of these galaxies do not seem to be directly connected with tidal perturbation and have
been discarded from our sample. We also removed from the sample galaxies such as NGC 5128 
(Cen A or Arp 153) because they belong to the category of polar ring ellipticals, whose gas 
content has already been discussed in a precedent study \citep {polar}. Similarly, all 
the galaxies containing gas- or star- counter-rotation have been removed. For 81 objects 
no distances have been found in the literature and for a further 486 galaxies no data are 
available on gas content. These galaxies have also been discarded. 
Among the 596 Vorontsov-Velyaminov galaxies we selected 395 systems only, while from the 
560 Arp and 6445 Arp \& Madore objects we selected 341 and 557 galaxies respectively. 
Finally, 27 more galaxies not included in the previous catalogues were added from 
\citet{zhu} work. Our final sample is composed of 1038 galaxies whose appearance seems to be 
perturbed by an external agent. The data on these galaxies are presented in Table ~1,
available in electronic form only. In the case of interacting pairs and triplets whose 
components appear detached, both components have been included. 

To analyze the ISM of these sample galaxies, we searched in the literature for survey 
papers \citep[e.g.][]{knapp, roberts, casoli, fabbiano, burstein, beuing, osullivan} 
and single studies in four different tracers: {\it warm dust} from 60 and 100 $\mu$m 
observations; {\it atomic gas} from 21 cm data; {\it molecular gas} from CO line observations 
and {\it X-ray luminosities} from Einstein and Rosat satellite observations. Many data come 
from LEDA\footnote{http://leda.univ-lyon1.fr/} galaxy catalogue \citep{leda} and from the 
NASA/IPAC extragalactic database (NED\footnote{http://nedwww.ipac.caltech.edu/}).
The list of sample galaxies has been compared with the \citet{veron2003} catalogue, to 
check the eventual presence of active galaxy nuclei (AGN). The number of galaxies
with data (detections and upper limits) and the number of detections only are listed 
in Table \ref{numbers}, binned according to the morphological type. 

To homogenize all these data, we reduced the published values to the distances extracted 
from a single source, the LEDA on-line archive.
The LEDA distance moduli are mainly derived from redshifts, corrected for 
Virgocentric inflow and adopting H$_o$=70 \kms\ Mpc$^{-1}$.  When redshift was not 
available, we used the photometric distance modulus, if present. For each galaxy 
we also collected the morphological classification, the diameter and the blue luminosity,
corrected for cosmological reddening and galactic- and internal- dust absorptions.
The blue corrected luminosity \LB\ in solar units and the square of the face-on corrected 
diameter in kpc Log D$^2$ were used as normalization factors for the gas masses and the X-ray 
luminosities. However, due to the similarity in the results obtained with one or 
other normalization factor, we present only the data normalized to \LB. 

The LEDA morphological type t codes derive from RC3 \citep{rc3} with some changes, as
visible in our Tables \ref{numbers} and \ref{values}. Type 5 includes systems with uncertain
classification (type S?).
In our sample 25 galaxies are unclassified in LEDA. We attributed the morphological type 
code t=11 to these systems in order to plot them in our graphs as well. 
We remember that the morphological classification may be misleading for very perturbed objects,
and when present in interacting galaxies reflects the present-day observed morphology and not 
the original one. The difference we found in the comparison with normal galaxies then will 
reflect the difference between the {\it present}, may be transient, morphology and that of a 
stable galaxy with similar appearance.

\subsection{Mass estimates}\label{methods}

We have estimated total gas masses and/or luminosities following standard procedures.
From the value of the IR fluxes at 60 and 100~$\mu$m we estimated the temperature 
(Kelvin) of the dust emitting at such a wavelengths:
\begin{equation}
T_d=49\ (S_{60}/S_{100})^{0.4}  
\end{equation}

and the corresponding warm dust mass, in solar units, by means of the expression: 
\begin{eqnarray}
Log\ M_d = -2.32\ +\ Log\ S_{100}+2\ Log\ d\ + \nonumber \\
   Log\ ( exp \left[ \frac{144.06}{T_d}\right] -1) 
\end{eqnarray}
In the above formula, $S_{100}$ is in mJy and $d$ in Mpc. 

HI gas masses were derived from 21 cm fluxes $S_{21}$ or from the m$_{21c}$ corrected
magnitudes of LEDA. From m$_{21c}$, we used the expression:
\begin{equation}
Log\ M_{HI} = 5.37-0.4(m_{21c}-17.4)+2\ Log\ d
\end{equation}
whereas if $S_{21}$, in Jy \kms, was available we used the formula:
\begin{equation}
Log\ M_{HI} = 5.37+\ Log\ S_{21}+2\ Log\ d 
\end{equation}
For 71 galaxies that are in close interaction or merging, the IR or 21 cm emission is 
attributed to the whole pair and not to a single galaxy in the literature. For instance, 
PGC 37967 and PGC 37969, making part of the object called `The Antennae', have no IR data 
from 12 to 100~$\mu$m visible in NED or LEDA catalogues. However, these data appear when the 
databases are searched with the name of the whole pair, ARP 224 or VV245. To manage these 
galaxies, we decided to calculate from the IR data the total warm dust mass of the pair. This 
total mass was normalized by the total absolute blue luminosity of the pair or the total area 
by adding the values of both components. So the values of both components of the pair appear 
identical in Tab.~1. These spatially unresolved data
were not used to define the mean values in the following statistical analysis, due to the 
difficulty in attributing the masses to every single object.

The formula used to derive molecular gas masses from CO(1--0) line fluxes ($S_{CO}$ 
in Jy \kms) is: 
\begin{equation}
Log\ M_{mol} = 4.17 + 2\ Log\ d + Log\ S_{CO}
\end{equation}
We implicitly assumed a constant CO/H$_2$ conversion factor
X = $N(H_2)/I_{CO}$= 2.3 10$^{20}$ mol cm$^{-2}$ (K km/s)$^{-1}$
\citep{strong}; $M_{mol}$ includes the helium mass fraction by multiplying the 
H$_2$ mass by 1.36. A possible variation of the CO-H$_2$ conversion factor X 
will be discussed later.

X-ray luminosities may derive from diffuse gas and/or from discrete sources. 
Very recent observations with the Chandra satellite indicate there is the possibility for 
some galaxies that X-ray fluxes are `contaminated' by the presence of X-ray 
binaries which must be taken into account in discussing the ISM physical 
properties \citep{kf}. We shall discuss the data according to the different 
models.

When no observed fluxes but only mass values were available in the consulted references, 
these have been scaled to the distances assumed here. If data from different sources are 
available for a galaxy, mean values have been calculated. When both upper limits and 
detections are available, only detections have been used to compute mean values. 
Moreover, if several upper limits exist, only the lowest value has been adopted.

In order to define the mean content of each tracer according to the morphological
type, we made use of survival analysis methods \citep[see][]{feigelson}. This method 
takes into account both detections and upper limits (UL) to derive representative 
averages. Our survival analysis was performed by means of the statistical package 
available in IRAF, using the Kaplan-Meier estimator. Mean values and errors listed
in Table \ref{values} derive from survival analysis of our data. When all the galaxies 
are detected, the survival analysis means coincide with arithmetic means. Conversely, 
the method cannot be applied when both upper
and lower limits are present, for instance when considering the mass ratios (e.g. 
\mmol/\mhi). In such a case, we present ratios of mean values. No survival analyses 
were made for unclassified galaxies (our type t=11). 

\begin{figure*}
\resizebox{18cm}{!}{\includegraphics[angle=0]{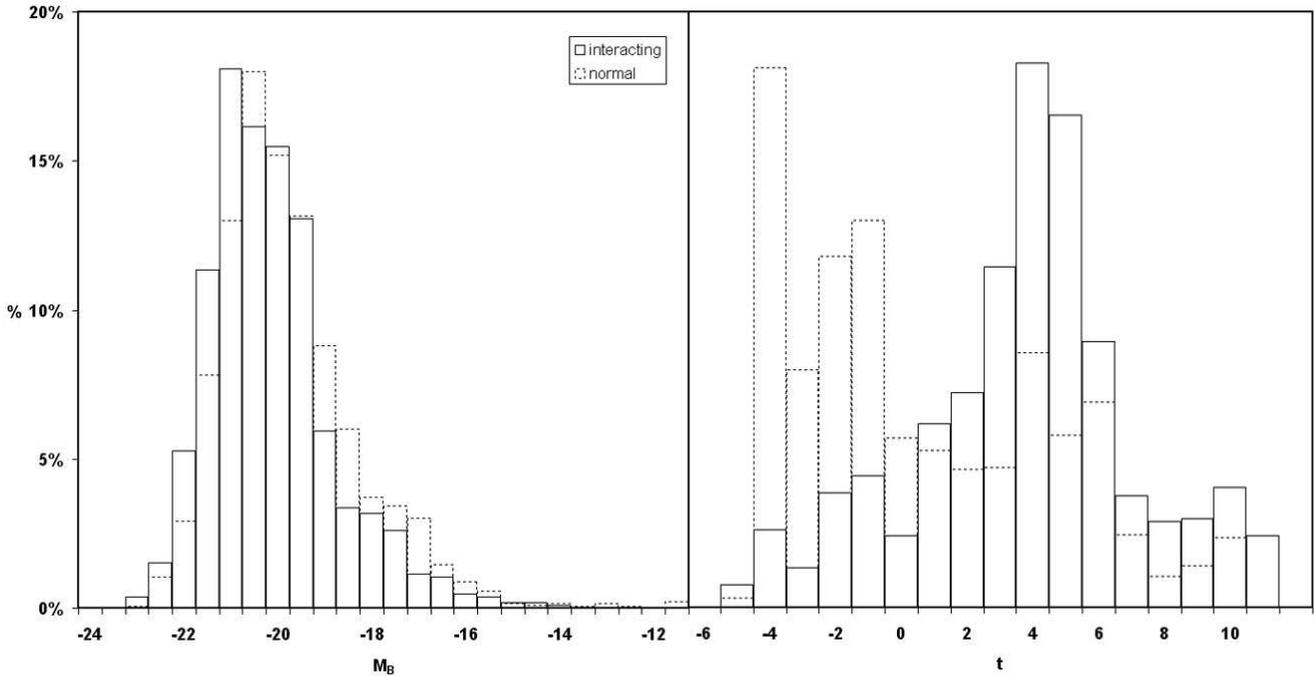}}
\caption{Histograms for absolute blue magnitude (left panel) and morphological type 
code (right panel) of interacting galaxies (full lines) compared to that of normal galaxies 
(dotted lines).} 
\label{isto1}
\end{figure*}

\subsection{Star formation}\label{sfr}
The production of stars inside strongly interacting systems may be enhanced during the 
close encounter or the merging of galaxies. Massive, newly born stars heat the dust
and increase the infrared luminosity \citep{lonsdale}.
Several authors estimated the star formation rate SFR, which is directly proportional to the
bolometric (40-100 $\mu$m) IR luminosity L$_{FIR}$ \citep{tronson,sage88,devereux91}, 
calculated from the fluxes S$_{60}$ and S$_{100}$. 
In LEDA catalogue, the entity of infrared emission is defined by the quantity m$_{FIR}$:

\begin{equation}
m_{FIR} = -2,5\ Log\ (2.58\  S_{60} + S_{100})+14.75
\end{equation}

We collected the m$_{FIR}$ values from LEDA catalogue for the sample of interacting galaxies
and for the comparison sample of normal galaxies of \citet{normal}. These values
give a direct estimate of the star formation in the two kinds of galaxies, when
corrected for the galaxy distance $d$, to give an absolute FIR luminosity 

\begin{equation}
L_{FIR} = 11.49 - 0.4\ m_{FIR}+ 2\ Log\ d
\end{equation}

FIR luminositiesa are plotted in Fig. \ref{LFIR} that is discussed in the next 
section.

Another parameter describing the stellar formation is the star formation efficiency, SFE,
indicating how much stars are produced by a unit of molecular mass. This value may be 
estimated dividing the FIR luminosity by the total mass of
molecular hydrogen: 
\begin{equation}
SFE = L_{FIR}/M(H_2)
\end{equation}

The more efficient a galaxy is in forming stars, the shorter the time for all the gas to 
be converted into stars, apart from possible recycling of mass through supernovae explosion or 
cooling flows \citep{sage88}. Several authors indicate that strongly interacting galaxies 
are also more efficient in forming stars \citep{sage88,combes}.
To study this effect, we plotted the values of Log(SFE) for both samples of 
galaxies in Figure \ref{SFE}. We note that the value of M(H$_2$) depends on 
two parameters: the observed CO line intensity and the assumed X conversion factor. 
Physically, X is determined by various factors, among which the metallicity and the 
temperature \citep[see][]{maloney,boselli2}. Generally galaxies with a lower metallicity have a X 
factor higher. However, we don't know the metallicity  of our sample galaxies
and our estimates of H$_2$ masses or SFE values depend from how much the assumed X value 
differs from the {\it true} X of each galaxy. In the literature, measured X values found 
in different type of galaxies span two order of magnitudes \citep[see][]{CO_H2}, being both
higher and lower than our assumed value. As consequence, our calculated H$_2$ masses will be 
lower than real if the {\it true} X value for each galaxy is higher than assumed, and higher than 
real if it is lower than the assumed value. The same happen for calculated SFE values. The 
spread of L$_{FIR}$/M(H$_2$)data visible in Figure \ref{SFE} may reflect both the effect of a 
different metallicity or temperature and that of a real different stellar formation efficiency.
\section{Properties of Interacting Galaxies}
The final data presented in Table ~1 are: (1) PGC number; (2) Other names of the galaxies,
mainly Arp, AM or VV; (3) morphological type code t; (4) Distance in Mpc; (5) log D, the logarithm 
of Diameter in Kpc; (6) log \LB, logarithm of the blue luminosity; (7) log \md/\LB, the dust 
content normalised to blue luminosity; (8) log \mhi/\LB, the HI content; (9) log \mmol/\LB, 
the molecular gas content; (10) log \Lx/\LB, the normalised X ray luminosity; (11) log \LFIR,
the FIR luminosity; (12) the presence of activity, extracted from \citet{veron2003} catalogue, 
indicated with an `a'; (13) the code for the references to the table values, as specified in 
the Appendix. Masses and luminosities are in solar units. 

\begin{table*}                                              
\caption{Mean Mass to light ratios of interacting galaxies for the different tracers, 
binned according the morphological types. For X-ray data, the corresponding luminosities 
are considered instead of the masses. All values are normalized to \LB, the corrected blue 
luminosities and are in solar units. The values have been obtained with survival analysis
technique.}                                              
\begin{center}
\begin{tabular}{lrrrrrc}
\hline                                              
\hline                                              
	&		& ~~~~~~Dust	& ~~~~~~~HI		& ~~~~~~Mol		& ~~~~~~X-ray		&  {\small $<$mol$>$/$<$HI$>$} \\
Type	&	t	& ~~Log M/L$_B$	& ~~Log M/L$_B$	& ~~Log M/L$_B$	& ~~Log L$_X$/L$_B$ &			\\
\hline                                              
E	&	-5	& -4.55$\pm$0.17	& -1.23$\pm$0.33	& ~~~~~~~~-~~~~	& -3.13$\pm$0.14	& ~~~~~~~-		\\
E$^+$	&	-4	& -4.90$\pm$0.28	& -1.53$\pm$0.52	& -2.70$\pm$~~-	& -2.91$\pm$0.18	& -1.16$\pm$0.52	\\
E/S0	&	-3	& -4.92$\pm$0.27	& -0.82$\pm$0.18	& -2.17$\pm$0.67	& -3.50$\pm$0.23	& -1.36$\pm$0.69	\\
S0	&	-2	& -4.19$\pm$0.13	& -1.11$\pm$0.24	& -2.24$\pm$0.30	& -3.25$\pm$0.27	& -1.14$\pm$0.38	\\
S0$^+$&	-1	& -3.70$\pm$0.12	& -0.81$\pm$0.12	& -0.68$\pm$0.02	& -3.37$\pm$0.38	&  0.13$\pm$0.12	\\
S0/a	&	0	& -3.83$\pm$0.08	& -0.68$\pm$0.14	& -0.94$\pm$0.13	& -3.23$\pm$~~-	& -0.26$\pm$0.19	\\
Sa	&	1	& -3.75$\pm$0.04	& -0.78$\pm$0.08	& -0.55$\pm$0.15	& -3.10$\pm$0.39	&  0.23$\pm$0.17	\\
Sab	&	2	& -3.76$\pm$0.05	& -0.81$\pm$0.06	& -0.71$\pm$0.15	& -3.72$\pm$0.25	&  0.09$\pm$0.17	\\
Sb	&	3	& -3.64$\pm$0.04	& -0.61$\pm$0.05	& -0.78$\pm$0.14	& -3.47$\pm$0.19	& -0.18$\pm$0.15	\\
Sbc	&	4	& -3.66$\pm$0.03	& -0.55$\pm$0.05	& -0.66$\pm$0.11	& -2.98$\pm$0.30	& -0.11$\pm$0.12	\\
Sc+S?	&	5	& -3.76$\pm$0.03	& -0.54$\pm$0.04	& -0.61$\pm$0.09	& -3.58$\pm$0.20	& -0.07$\pm$0.10	\\
Sc	&	6	& -3.71$\pm$0.08	& -0.39$\pm$0.06	& -1.19$\pm$0.19	& -4.43$\pm$0.30	& -0.81$\pm$0.20	\\
Scd	&	7	& -3.88$\pm$0.07	& -0.52$\pm$0.08	& -1.07$\pm$0.25	& -3.52$\pm$0.27	& -0.55$\pm$0.26	\\
Sd	&	8	& -3.94$\pm$0.09	& -0.22$\pm$0.12	& -1.41$\pm$0.31	& -3.56$\pm$0.05	& -1.19$\pm$0.33	\\
Sm	&	9	& -3.94$\pm$0.14	& -0.25$\pm$0.10	& -1.78$\pm$0.37	& -3.60$\pm$0.28	& -1.53$\pm$0.38	\\
Irr	&	10	& -4.07$\pm$0.09	& -0.28$\pm$0.10	& -1.07$\pm$0.32	& -2.61$\pm$0.63	& -0.79$\pm$0.34	\\
\hline                                              
\hline                                              
\end{tabular}                                              
\end{center}
\label{values}                                              
\end{table*}                                              

In the following discussion we compare the mean properties of different gas tracers in our 
sample with that for a sample of 1916 normal galaxies published by \citet{normal}. 
The distributions of galaxies in the two samples, in function of absolute magnitudes
and morphological types, are shown in Fig. \ref{isto1}. Among the interacting galaxies considered 
here, the percentage of early-type galaxies is very small, the 15.4\% of the total, while 
the sample of normal galaxies has half of systems with t$<$0 (see Fig. \ref{isto1}). 
However, most part of our comparison will be made separating the galaxies according to 
the morphological type code, so keeping into account this difference. A Kolgomorov-Smirmov 
(KS) test performed on the luminosity and linear diameter functions shows that N(M$_B$) and 
N(Log D$^2$) are significatively different in the two samples. The same test indicates that the 
distributions are statistically similar if the interacting galaxies histograms are shifted 
by $\Delta$M$_B$=0.5  and $\Delta$(log D$^2$)=0.1 with respect to that of normal galaxies. 
In other words, interacting galaxies appear brighter by 0.5 mag and larger by the 25\% 
with respect to the normal galaxies. Both differences may derive from the effects of the 
interactions. 

\begin{figure}
\resizebox{9cm}{!}{\includegraphics[angle=0]{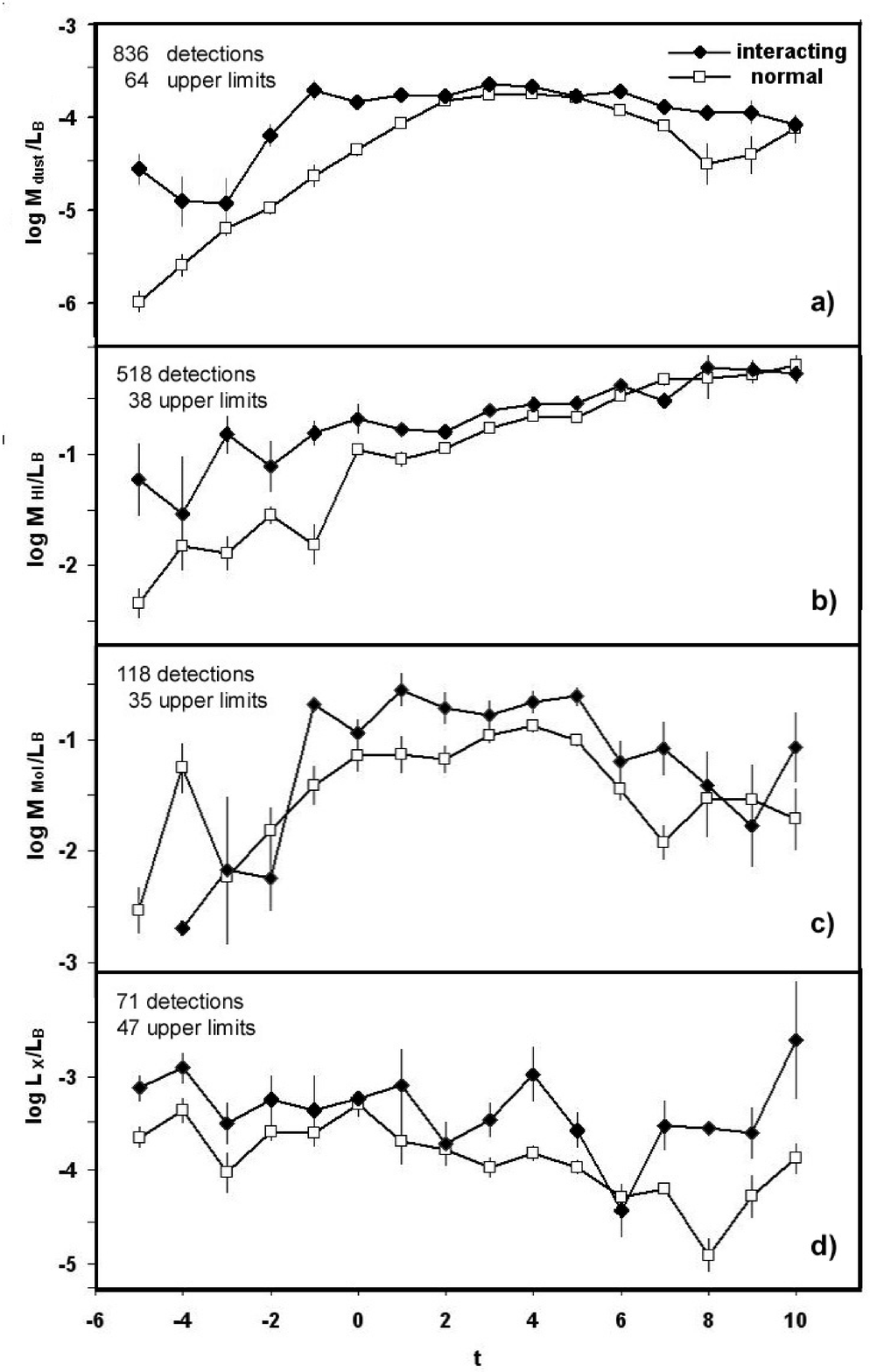}}
\caption{Comparison between mean values from Table \ref{stat} found for interacting galaxies and
corresponding means for normal galaxies \citep{normal}. Bars represent 1 $\sigma$ errors.} 
\label{survival}
\end{figure}

The mean values of gas content obtained with survival analysis for each tracer (dust, HI,
molecular gas, X-ray) and for each morphological type are presented in Table \ref{values}
and Fig. \ref{survival}. We tested the statistical significance of the differences found 
between interacting and normal galaxies samples applying a T-test to the mean values and 
an F-Test to the variances. The significance of the single values has also been checked by 
a $\chi^2$ test, assuming the values of normal galaxies as expected values. The results of all 
these tests are presented in Table \ref{stat}. 

Due to the different behaviour of galaxies with different t, we decided do not make the 
comparison between the two samples mixing together galaxies of all morphological types, as 
sometimes seen in the literature. For statistical tests only, we separated galaxies in two 
large categories that present similar properties in all the tracers, as explained in the next 
sections: early types (-5 $\ge$ t $\ge$0) and late types (1$\ge$t$\ge$10).

\subsection{Dust and cold gas}
The FIR fluxes at 60 and 100 $\mu$m trace the warm dust content, at temperatures around 
40 K. Most of the dust in galaxies is however colder, and can be studied analyzing 
the data from longer wavelength, from 100 to 200 $\mu$m \citep{popescu}. Our 
sample galaxies possess data essentially from IRAS observations, so we shall discuss
their dust mass keeping in mind that the total values may be one order of magnitude 
greater. Given that our comparison with normal galaxies is made using similar data, our 
conclusions about the warm dust content should be consistent.

Looking at the Log \md/\LB\ values obtained with survival analysis for each morphological 
type (Fig. \ref{survival}a), one can see that the warm dust content is 
quite similar to that of normal galaxies for types later than t=2 (Sab) but higher for early 
type galaxies. Early type interacting galaxies have {\it in mean} a warm dust content one order 
of magnitude higher than normal galaxies of the same type. According to statistical tests
(Table \ref{stat}) the significance of this difference is high: the two samples of early type 
galaxies have mean  values that differ by more than 95\% of significance. They show however a 
similar dispersion of values (F-test). Late types show on the contrary a lower difference between 
normal and interacting galaxies and the similarity in dust content is stronger if the comparison 
is restricted to type from Sab to Sd, excluding Irregular galaxies (see the third line of 
Table \ref{stat}).

This excess of mass content seen in early type galaxies is also evident for atomic gas 
(log \mhi/\LB,  see Fig. \ref{survival}b) but does not seem to be present as far as the 
molecular gas content is considered (log \mmol/\LB, see Fig. \ref{survival}c). The statistical
tests show a significant difference for HI and a strong similarity for molecular gas. Ellipticals
and S0 galaxies (t$<$-1) have low values of Log \mmol/\LB\  with respect to the sequence 
of spiral galaxies, but are based in total on 10 detections only, compared to 71 detections for 
early type systems in the sample of normal galaxies. 

The mean Log \mhi/\LB\ for late type interacting galaxies appears similar to that of normal
ones to an higher statistical significance. For these types, the differences between 
interacting and normal galaxies appear when log \mmol/\LB\  is considered. The two samples
show the same dispersion of values but mean molecular gas contents significantly 
different, being interacting galaxies richer than normal ones. 

In the literature, \citet{horellou2} find that the HI surface  density in interacting 
galaxies, mediated over all the morphological types, is equal to 6.33$\pm$ 0.51 \Msun\  
Kpc$^{-2}$. This value is lower of that found by  \citet{HG} for `isolated' galaxies, 
corresponding to 6.81$\pm$0.24. In this latter sample most part of detected galaxies are 
of morphological type from Sa to later. On the contrary, our values of Log \mhi/\area\ 
show that early type interacting galaxies have an higher HI content per Kpc$^2$ than normal, 
as seen for Log \mhi/\LB\  curves plotted in Fig. \ref{survival}b, while the values are similar 
for late-type. If we mediate our values from survival analysis over the morphological types 
we obtain a mean value of Log \mhi/\area= 6.7 $\pm$0.3, based on 518 detections and 38 upper
limits. This value increases to 6.9 $\pm$ 0.2 if types from Sa to Irr only are considered. On the 
contrary, mediating all the  Log \mhi/\area\ values published by \citet{normal} and based on 891 
detections and 220 upper limits, we find a mean Log \mhi/\area\ value of 6.5 $\pm$ 0.62 that 
becomes 6.8 $\pm$0.2 if late type galaxies only are considered. Both values of normal
galaxies are consistent with that found by \citet{HG} in their sample of 288 detected galaxies. 

\begin{figure}
\resizebox{9cm}{!}{\includegraphics[angle=0]{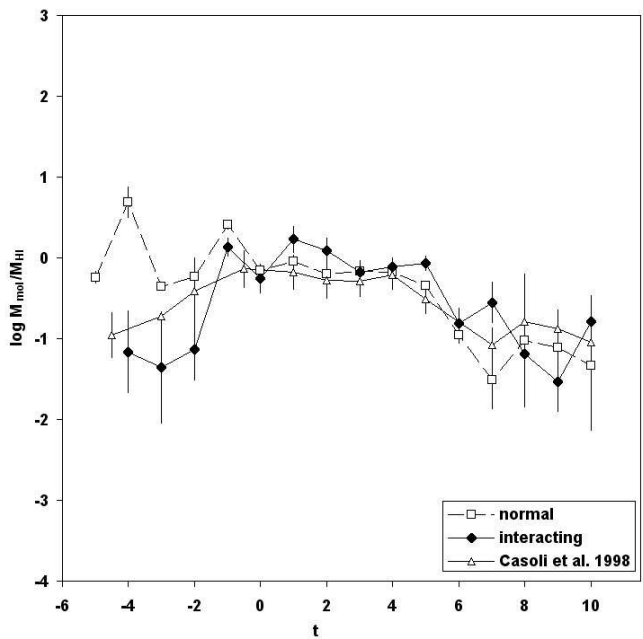}} 
\caption{The ratio between molecular and atomic gas according to the morphological type t.
Mean ratios for interacting galaxies are compared with that of normal galaxies and with the
`mixed' sample of \citet{casoli}. } 
\label{mol_HI}
\end{figure}

The change of the molecular/HI gas ratio with morphological type has been studied 
calculating (log $<$\mmol$>$/$<$\mhi$>$) from the mean values of log \mmol/\LB\  and 
log \mhi/\LB\ of Table \ref{values}. The resulting mol/HI ratio appears very low for 
early-type systems, if compared with those of normal galaxies \citep{normal} and similar 
to them for types later than -2. Similar low values for early type galaxies were found by 
\citet{casoli}. Looking at the papers cited as sources of their data \citep[see in][]{casoli}, 
we found that in their sample 126 galaxies out of 456 (28\%) are peculiar, having disrupted 
structures (e.g. NGC 520) or polar rings (e.g. NGC 2685) or counter-rotations, all phenomena 
typical of interaction with the environment. This means that the \citet{casoli} values
are contaminated by the presence of these peculiar galaxies and that early-type interacting 
systems really have lower mol/HI values with respect to non-interacting, normal galaxies.

Resuming the above considerations, we find a difference in the colder component of those 
interacting galaxies classified early-type (t$<$0) with respect to those classified as late-type 
(t$>$0): the first ones have dust and HI overabundance, even with respect to their molecular gas, 
while the second are richer in molecular gas and are quite normal, on average, for dust and 
atomic gas content. This behavior may be due to the characteristics of the interaction and the 
subsequent evolution of the gas, as discussed later. 

The fact that interacting galaxies have a luminosity function brighter than normal ones,
as visisble in Fig. \ref{isto1}, strengthen our conclusion on gas overabundances. In fact, the adopted
normalisation factor \LB\ should reduce the log M/\LB\ for interacting galaxies, contrary to 
what observed.  

\begin{figure}
\resizebox{9cm}{!}{\includegraphics[angle=0]{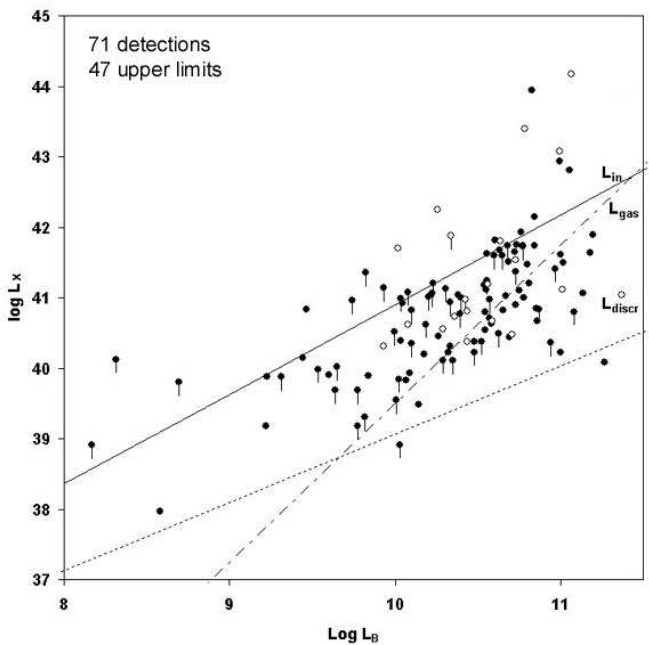}} 
\caption{The X-ray luminosity (W m$^{-2}$) versus the blue luminosity (solar units) for
interacting galaxies. Active galaxies are represented as open circles. A bar under the circle
indicates an upper limit. The lines represent the values expected from hot discrete sources 
(dotted line) or hot diffuse gas (dashed-dotted line). 
The full line is the limit expected from steady-state cooling flows (see text for details). Most 
of detected sources lie near the line of hot diffuse gas.} 
\label{LXLB}
\end{figure}

\subsection{X-ray emission}
To study the component at higher energy of these galaxies, we analyzed the properties of the 
X-ray emission and the presence of Active Galaxy Nuclei by the \citet{veron2003} catalogue. 
Among our sample galaxies, 71 or 7\% are classified as AGN. This percentage is similar to 
that found for the sample of normal galaxies \citep{normal}, i.e. 8\%, so in this respect the 
strongly interacting galaxies do not appear to be different from normal ones. In Figure 
\ref{survival} we note that the survival analysis mean values are always higher than that of 
normal galaxies (including AGN). It is clear that many AGN have high values of Log \Lx/\LB, but 
looking at the single points, one can see that the rest of interacting galaxies also has higher 
log \Lx/\LB\ values than normal galaxies, independent from the morphological type. 

To understand the origin of the emission, we plotted the log \Lx\  of the single galaxies
versus log \LB\ and we compared the observed values with theoretical expectations. In the plot 
the representative points of galaxies hosting nuclear activity are distinguished by open symbols. 
In Figure \ref{LXLB} the lines corresponding to the X-ray emission foreseen for diffuse gas 
(labeled L$_{gas}$) and from discrete sources inside the galaxy (L$_{discr}$) are plotted. 
In the first case, the X-ray luminosity would be proportional to the square of blue luminosity
\LB\  while in the second one it would be directly proportional to \LB\  \citep{ciotti}. 
Most of our non-AGN galaxies are clustered around the line expected from emission of diffuse 
gas. Their luminosity is lower than the values expected for cooling flows (line labeled L$_{gas}$)
assuming a SN rate of 0.18 \citep{cappellaro}. Both these aspects are consistent with
the hypothesis that this X-ray flux derives from gas stripped from the galaxies by ram pressure 
during their motion or, alternatively, from gas emitted by SN winds in the intergalactic space.

The origin of this gas has been studied in detail in the literature for some of our systems:
\citet{davis} found a diffuse X-ray emitting gas confined near the center of the NGC 2300 group 
(Arp 114) in addition to point sources localized on the single galaxies. The presence of 
diffuse gas in a common potential well is a feature visible in other galaxy pairs \citep{henriksen} 
but some cases (e.g. Arp 116 = NGC 4647/49) have X-ray emission coming from a single member 
of the pair only. Given the various X-ray morphologies, a common model in the origin of the X-ray 
emitting gas is difficult to apply to all our interacting galaxies. 

\begin{table}                                              
\caption{Parameters of the statistical tests applied to our sample galaxies in comparison with
normal ones (see text). Columns labeled P() represents the probability that the observed differences
are statistically significants.}                                              
\tabcolsep 0.15truecm
\begin{center}
\begin{tabular}{llrrrrr}
\hline                                              
Tracer &	Type	& $\chi^2$	&	k	& P($\chi^2$)&	P(T-Test)	&	P(F-Test)	\\
\hline															
\hline															
Dust	&	early	&	13.39	&	5	&	98.0\% &	95.9\%	&	24.2\%     \\
Dust	&	late 	&	8.29	&	9	&	49.5\% & 	94.9\%	&	93.4\%	\\
Dust	&    Sab-Sd  &	-	&	-	&	- 	 &	84.6\%	&	95.6\%	\\
\hline															
HI	&	early	&	16.58	&	5	&	99.5\% &	98.7\%	&	54.6\%	\\
HI	&	late 	&	5.25	&	9	&	18.8\% &	49.3\%	&	67.5\%	\\
\hline															
mol	&	early	&	19.47	&	4	&	99.9\% &	2.7\%		&	65.4\%	\\
mol	&	late 	&	21.34	&	9	&	98.9\% &	94.5\%	&	29.2\%	\\
\hline															
X-ray	&	early	&	14.95	&	5	&	99.0\% &	97.6\%	&	37.3\%	\\
X-ray	&	late 	&	47.26	&	9	&	100.0\% &	99.5\%	&	63.3\%	\\
\hline															
\hline														
\end{tabular}                                              
\end{center}
\label{stat}                                              
\end{table}                                              

\subsection{Star formation}
\begin{figure}
\resizebox{9cm}{!}{\includegraphics[angle=0]{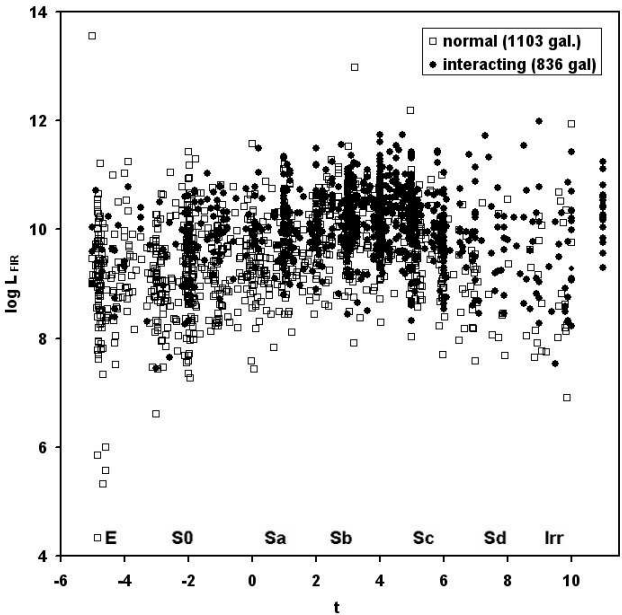}} 
\caption{The absolute far infrared luminosities vs. morphological type code of the single 
interacting galaxies (full circles) compared to th single normal galaxies include in the sample 
of \citet{normal}.} 
\label{LFIR}
\end{figure}

\begin{figure}
\resizebox{9cm}{!}{\includegraphics[angle=0]{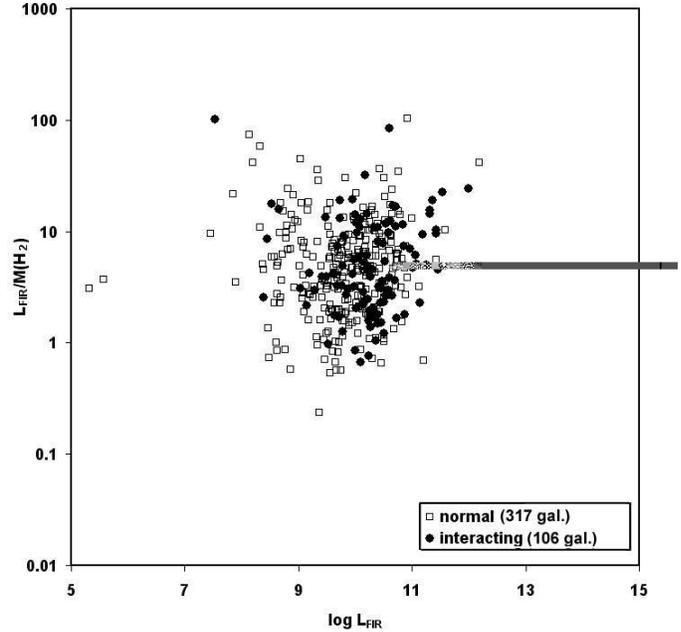}} 
\caption{The star formation efficiency per unit of molecular gas mass, indicated by the ratio 
L$_{FIR}$/M(H$_2$), plotted vs. the far infrared luminosity. Interacting and normal galaxies 
appear to have similar ranges of values, even if interacting galaxies are more luminous in the 
infrared.} 
\label{SFE}
\end{figure}

In spiral galaxies (and probably only in part in early-types galaxies) the FIR luminosity is 
connected with the massive star formation rate \citep{devereux97}.
An enhanced stellar formation in interacting galaxies, already known in the literature, is also 
visible in Figures \ref{LFIR} and \ref{SFE}. 

In the first one, interacting galaxies often appear
more luminous in infrared than normal ones, but with a large spread of values. 
High FIR luminosity is evident in those systems with no morphological classification (right edge 
of the Figure). These latter are often systems with disrupted morphology, probably in a phase of 
strong merging, that appear to also be actively star-forming. 
We must note, however, that also normal galaxies (i.e. non-perturbed,
with no polar ring or known kinematical decoupling between gas and stars) may exhibit high values
of star formation rate, even those classified as ellipticals. 

The star formation efficiency (Fig. \ref{SFE}) appears with a similar dispersion of values.
Despite a higher FIR luminosity and star formation rate, the interacting galaxies however 
do not seem to be more efficient in forming stars. The two distributions intersect together but 
with a clear deviation of interacting systems toward higher levels of L$_{FIR}$. This effect is
different from that found by \citet{sage88,horellou3} and by \citet{combes} which suggest, for 
interacting pairs, an higher efficiency in forming stars than pairs with larger separation. 
However, \citet{combes} find that in their sample of paired galaxies log L$_{FIR}$/M(H$_2$) 
does not change in function of the separation between components. 
The range of SFE values of our galaxies is larger than that of the previous authors 
and covers the differences found by them. This may explain our lack of differences between our 
samples of interacting and normal galaxies. In our data, even binning the interacting galaxies 
in groups of different separation (i.e. merging,
semidetached components, separated components with tails or bridges, single galaxies), the plot of 
Figure \ref{SFE} does not change and the groups have representative points mixed together. At the 
same time, there are no apparent dependences from the morphological type t, including 
the unclassified objects (t=11). 

\section{Conclusions}
The category of galaxies considered in this paper is a zoo containing several types of
objects, going from galaxies observed during a close encounter (e.g. M51) to merging systems 
(e.g The Antennae) or galaxies relatively isolated but still perturbed after a 
past encounter (e.g. M82). The spread of values found here and in the literature may derives
from the fact that different kind of perturbations generate different effects in galaxies.

We find that the morphological type also have a role in determining the properties
of these stellar systems and that may be not correct to mix together interacting galaxies of
all morphological types. 

In addition, the comparison with the literature is complicated by the fact that
very often authors use as `normal' galaxies a list of objects that contains known cases of
peculiar galaxies. For instance, in Table 3 of \citet{devereux97} is visible NGC 660, a 
disrupted polar ring galaxy. \citet{HG} include several Arp and VV galaxies while \citet{casoli} 
present a sample containing many dynamically peculiar galaxies. The data of these papers
have been used by other authors to deduce properties of peculiar galaxies in comparison with
`normal' ones. The sample of normal galaxies by \citet{normal} used here may also contains 
galaxies that in the future may be discovered as 'peculiar', but have been already cleaned 
by all the known cases of interacting and dynamically peculiar galaxies known. 

From our data, we see that interacting or disturbed galaxies have always more gas than normal 
ones, and that this gas is mainly in the form of atomic hydrogen in early type systems, while it 
is in  molecular form for galaxies of latest types. All have a higher X-ray luminosity.
Because of these differences, we shall discuss the early type systems separately from the spiral 
ones, comparing our finding with that from literature.

A galaxy that appears now as an elliptical may be in the final stage of a merging process,
after the complete fusion of the stellar content of two galaxies. Alternatively, it may be
a galaxy deprived of the gas because of a close encounter with a more massive one. In both 
processes, one may expect that the pre-existing gas in the two systems would be heated,  
with the conversion of molecular gas in atomic form and by the creation of
an X-ray component. This scenario agrees with the higher HI and X-ray content found
in interacting early type galaxies. 

A more complicated scenario appears from the data concerning galaxies classified as spirals.
In these systems the mean HI and dust content seems to be similar to that of an unperturbed,
normal galaxy of the same morphological type. In addition to the excess in X-ray luminosity,
these galaxies seem to have a molecular gas content always one order of magnitude higher
that that of normal galaxies. There are two possible explanation for this effect.

\begin{figure}
\resizebox{9cm}{!}{\includegraphics[angle=0]{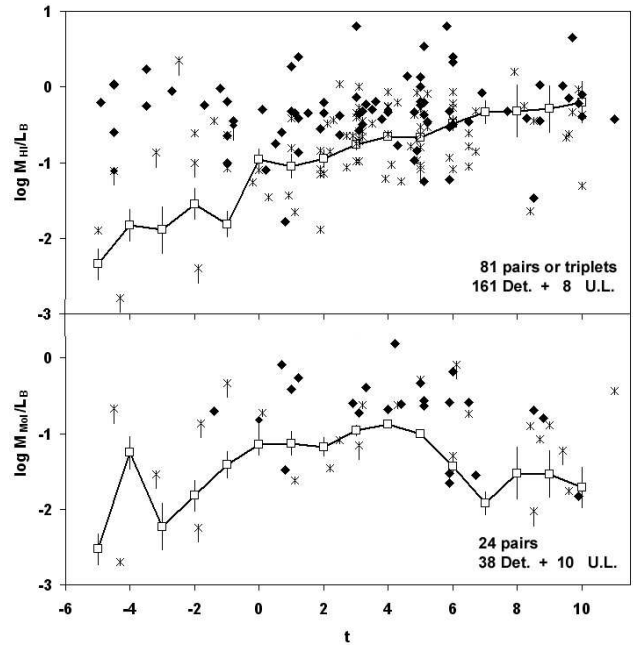}} 
\caption{Distribution of the representative points for 179 galaxies in pairs or triplets 
present in our sample for which detections or upper limits are available. The higher mass member 
(primary) is shown with a full diamond while the lower mass ones (secondary) are indicated with 
a cross. } 
\label{coppie}
\end{figure}

The first possibility is that during the collision the galaxies stimulate a gas inflow 
by means of gravitational torques, that enhance the CO luminosity  because of a massive
accumulation of molecular gas. This hypothesis has been presented by \citet{combes}
to explain the observed characteristics of a sample of 51 paired galaxies, whose interacting
subsample appears more luminous in CO line than the remaining binaries with higher separation.
We tested the idea that the gas is simply exchanged by the two components of the same
pair, enriching one of them and depauperating the other. We call these galaxies primary and 
secondary respectively. In our sample there are 179 galaxies in interacting pairs or triplets 
where the single galaxies are separated and have literature data at 21cm or CO lines. Plotting 
in Figure \ref{coppie} (upper panel) the values of log \mhi/\LB\ of these galaxies only and 
identifying with different symbols the primary and the secondary member, one can see that spirals 
which are primary members have -on mean- higher values than normal galaxies of same morphological 
type. On the contrary, the secondary members, of whatever morphological type, have values generally 
lower, below that of normal galaxies. This is in agreement with the idea that part of atomic gas 
has been transferred between them. However, this is not true when log \mmol/\LB\ values are 
considered (Figure \ref{coppie}, lower panel): in this case, the primary members of the pair show 
an excess of molecular mass and the secondary members have in general values near or over the mean 
curve of normal galaxies. We may suppose that less dense, atomic gas, may be exchanged during an 
encounter more easily than the molecular gas, condensed in clouds. The existence of a molecular 
mass excess in these paired galaxies is then difficult to explain supposing a simple transfer 
during the close encounter. 

The second, completely different possibility is that the molecular gas excess is not real but due 
to different physical conditions inside the galaxy. In the totality of cases presented here 
the \mmol\ is determined from CO line luminosities, transformed in solar masses adopting a X 
conversion factor, as explained in Section \ref{methods}. This value depends on the metallicities 
or on the physical conditions of the molecular gas and in different kind of galaxies may assume a wide 
range of values (see the discussion in Section \ref{sfr}). An overestimate of the molecular mass 
may derives by the assumption of a X factor lower than real. among the various possibility, the 
presence of gas of high temperature may overestimate the H$_2$ mass. However, in some well known 
cases present in our sample, such as the already cited `Antennae' or `Atoms for Peace' (PGC 68612), 
believed to be the result of a two spirals merger, the H$_2$ mass excess is accompanied by an HI 
mass excess. Similarly the `Whirlpool galaxy' M51 and its satellite (PGC 47404 and PGC 47413), show 
an excess of gas mass. In cases like these, we may think that the gas excess is real for the 
molecular component like it is for the atomic one. 

Another suggestion on the reality of the higher molecular masses comes from the analysis of
the star formation rates and efficiencies: interacting galaxies appear more luminous in the 
infrared (Fig. \ref{LFIR}) because of the higher number of newly formed stars heating the dust 
\citep{tronson}. This higher star formation rate does not correspond to a different efficiency in 
star formation per unit of calculated H$_2$ mass. This may be explained if the higher star formation 
is simply due to a higher molecular gas quantity. Higher H$_2$ mass and concentration in perturbed 
galaxies was already suggested by \citet{braine}. In our data, the large spread of data in Figure 
\ref{SFE} may indicate that the X conversion factor is different between galaxies but even in this 
hypothesis, the range of variation inside interacting galaxies does not appear different from that 
of normal, non interacting stellar systems. Our conclusion are different with the higher SFE 
found in interacting galaxies by previous authors \citep{sage88,combes,horellou3}. As discussed 
before, their comparison sample contains peculiar galaxies and the large spread of values found 
in our sample cover the differences found by them. We must conclude that, with the present data, 
our sample of interacting galaxies appears richer in molecular gas and in some cases also in atomic 
gas. 

It is not clear from where this mass excess originates. An excess of gas, both atomic and 
molecular, of one order of magnitude is known in the literature for another class of peculiar 
objects: the polar ring galaxies \citep{polar}. In that case, the matter forming the ring is 
accreted from outside. May this also be the case of interacting galaxies?


\begin{acknowledgements}
The authors thanks Dr. C. Horellou for her useful suggestions during the
revision of this paper. 
This research made use of LEDA database, the Vizier service \citep{vizier}, the 
SIMBAD database (operated at CDS, Strasbourg, France), the NASA/IPAC 
Extragalactic Database (NED) (which is operated by JPL, California 
Institute of Technology, under contract to NASA) and 
NASA's Astrophysics Data System Abstract Service (mirrored in CDS of
Strasbourg). GG made use of funds from the University of Padova (Fondi 60\%-2003).
\end{acknowledgements}


\section{Appendix A: References to Table 2.  }
{\small
\begin{tabular}{lll}
\hline
code & Source of data & ISM \\
  &                    & tracer \\
\hline
 1a  &  \citet{andreani}  &  HI   \\
 1b  &  \citet{andreani}  &  CO  \\
 2  &  \citet{beuing}     &  X-ray  \\
 3  &  \citet{bregman}    &  CO  \\
 4  &  \citet{burstein}   &  X-ray  \\
 5a  &  \citet{casoli91}  &  HI  \\
 5b  &  \citet{casoli91}  &  CO  \\
 6  &  \citet{casoli}     &  CO  \\
 7  &  \citet{fabbiano}   &  X-ray  \\
 8a  &  \citet{horellou}  &  HI  \\
 8b  &  \citet{horellou}  &  CO  \\
 9  &  \citet{knapp}      &  IR  \\
10a  &  \citet{lavezzi}   &  IR  \\
10b  &  \citet{lavezzi}   &  HI  \\
10c  &  \citet{lavezzi}   &  CO  \\
11a  &  LEDA \citep{leda} &  IR  \\
11b  &  LEDA \citep{leda} &  HI  \\
12a  &  NASA/IPAC Extragalactic Database (NED)      &  IR  \\
12b  &  NASA/IPAC Extragalactic Database (NED)      &  HI  \\
12c  &  NASA/IPAC Extragalactic Database (NED)      &  X-ray  \\
13   &  \citet{osullivan}      &  X-ray  \\
14a  &  \citet{roberts}   &  IR  \\
14b  &  \citet{roberts}   &  HI  \\
14c  &  \citet{roberts}   &  CO  \\
14d  &  \citet{roberts}   &  X-ray  \\
15a  &  Sage 1993         &  HI  \\
15b  &  Sage 1993         &  CO  \\
16  &  \citet{vandriel}   &  HI  \\
17a  &  \citet{welch}    &  HI  \\
17b  &  \citet{welch}    &  CO  \\
18  &  \citet{wiklind}    &  CO  \\
19 &  \citet{fcrao}      &  CO  \\
20 &  \citet{zhu}      &  CO  \\
\hline
\end{tabular}
}
\noindent
%

\noindent

%
\noindent


\begin{thebibliography}{}
\bibitem[Andreani, Casoli \& Gerin(1995)]{andreani} Andreani, P., Casoli, 
	F.\ \& Gerin, M.\  	1995, A\&A, 300, 43 
\bibitem[Arp(1966)]{arp} Arp, H.\ 1966, Pasadena: California Inst.\ 
	Technology, 1966,  
\bibitem[Arp \& Madore(1987)]{am} Arp, H.\ C.\ \& Madore, B.\ F.\ 1987, 
	Cambridge Univ.\ Press, 1 (1987), 1 
\bibitem[Athanassoula(2002)]{lia} Athanassoula, E.\ 2002, ApSS, 281, 39 
\bibitem[Bendo \& Barnes(2000)]{bendo} Bendo, G.~J.~\& Barnes, J.~E.\ 2000, 
     MNRAS, 316, 315 
\bibitem[Bertola \& Galletta (1978)]{bg}Bertola, F., Galletta, G., 1978, 
	ApJ 226, L115
\bibitem[Bettoni \etal (2001)]{polar} Bettoni, D., Galletta, G., 
	Garc\'{\i}a-Burillo, S. Rodr\'{\i}guez-Franco, A.,2001, A\&A, 374, 421 
\bibitem[Bettoni \etal (2003)]{normal}Bettoni, D., Galletta, 
G., Garc{\'{\i}}a-Burillo, S. 2003, A\&A, 405, 5 
\bibitem[Beuing \etal(1999)]{beuing} Beuing, J., D\"obereiner, S., 
	B\"ohringer, H., Bender, R., 1999, MNRAS, 302, 209
\bibitem[Boselli(2001)]{CO_H2} Boselli, A.\ 2001, ESA SP-460: 
The Promise of the Herschel Space Observatory, 357 
\bibitem[Boselli et al.(1995)]{boselli} Boselli, A., Gavazzi, G., Lequeux, 
	J., Buat, V., Casoli, F., Dickey, J.\ \& Donas, J.\ 1995, A\&A, 300, L13 
\bibitem[Boselli \etal (2002)]{boselli2} Boselli, A., Lequeux, J., \& 
         Gavazzi, G.\ 2002, ApSS, 281, 127 
\bibitem[Bregman \etal(1992)]{bregman} Bregman, J.N., Hogg, D.E., 
	Roberts,M.S., 1992, ApJ, 387, 484
\bibitem[Braine \& Combes(1993)]{braine} Braine, J., Combes, F., 1993, A\&A, 269, 7
\bibitem[Burstein \etal(1997)]{burstein} Burstein, D., Jones, C., Forman, 
	W., Marston, A.P., Marzke, R.O., 1997, ApJS, 111, 163
\bibitem[Cappellaro, Evans, \& Turatto(1999)]{cappellaro}Cappellaro, E., Evans, R., 
          \& Turatto, M.\ 1999, A\&A, 351, 459 
\bibitem[Casoli \etal (1991)]{casoli91} Casoli, F., Boiss\'e, P., Combes, F.\ 
           \& Dupraz, C.\ 1991, A\&A, 249, 359 
\bibitem[Casoli \etal (1998)]{casoli} Casoli, F., Sauty, S., Gerin, M., Boselli, A., 
       Fouqu\'e, P., Braine, J., Gavazzi, G., Lequeux, J., Dickey, J. 1998, A\&A, 331, 451
\bibitem[Ciotti \etal (1991)]{ciotti} Ciotti, L., 
       Pellegrini, S., Renzini, A., \& D'Ercole, A.\ 1991, ApJ, 376, 380 
\bibitem[Corsini \& Bertola(1998)]{crbis} Corsini, E.~M.~\& Bertola, F.\ 1998, Journal 
	of Korean Physical Society, 33, 574 
\bibitem[Combes  \etal (1994)]{combes} Combes, F., Prugniel, P., Rampazzo, R., 
           Sulentic, J.W.\ 1994, A\&A, 281, 725 
\bibitem[Davis \etal (1996)]{davis} Davis, D.~S.~ Mulchaey, J.~S., Mushotzky, R.~F.,
       Burstein, D.,\ 1996, ApJ, 460, 601 
\bibitem[de Vaucouleurs \etal (1991)]{rc3} de Vaucouleurs G., de Vaucouleurs
        A., Corwin H.G., Buta R.J., Paturel G., Fouque P., 1991, Third 
	Reference Catalogue of Bright Galaxies (RC3), Springer-Verlag: 
	New York
\bibitem[Devereux \& Young(1991)]{devereux91} Devereux, N.~A.~\& Young, J.~S.\ 1991, 
      ApJ, 371, 515 
\bibitem[Devereux \& Hameed (1997)]{devereux97} Devereux, N.~A.~\& Hameed, S.\ 1997, 
      AJ, 113, 599 
\bibitem[Dressler et al.(1997)]{dressler97} Dressler, A.~et al.\  
        1997, ApJ, 490, 577 
\bibitem[Fabbiano \etal (1992)]{fabbiano} Fabbiano, G., Kim, D.-W., 
	Trinchieri, G., 1992, ApJS, 80, 531
\bibitem[Feigelson \& Nelson(1985)]{feigelson} Feigelson, E.D., \& Nelson, P.I.,
	1985, ApJ, 293, 192 
\bibitem[Galletta(1996)]{crgg} Galletta, G., 1996, in Barred Galaxies, IAU
      Coll 117, ed Buta R., Crocker D.A. Elmegreen B.G., ASP Conf. Ser. 
	91, 429
\bibitem[Gerin \& Casoli(1994)]{gerin} Gerin, M.\ \& Casoli, F.\ 1994, 
	A\&A, 290, 49 
\bibitem[Haynes \& Giovanelli(1984)]{HG} Haynes, M.~P.~\& Giovanelli, R.\ 1984, AJ, 89, 758 
\bibitem[Henriksen \& Cousineau(1999)]{henriksen} Henriksen, M.~\& Cousineau, 
         S.\ 1999, ApJ, 511, 595 
\bibitem[Horellou, Casoli \& Dupraz(1995)]{horellou} Horellou, C., Casoli, 
	F.\ \& Dupraz, C.\ 	1995, A\&A, 303, 361 
\bibitem[Horellou \& Booth(1997)]{horellou2} Horellou, C., Booth, R.,S.  1997, A\&AS, 126, 3 
\bibitem[Horellou \etal(1999)]{horellou3} Horellou, C., Booth, R., S.~\& Karlsson, B. 1999, ApSS, 
   269, 629
\bibitem[Kim \& Fabbiano(2003)]{kf} Kim, D.~\& Fabbiano, G.\ 2003, ApJ, 586, 826 
\bibitem[Knapp \etal(1989)]{knapp}Knapp, G.R., Guhathakurta, P., Kim, D.-W.,
       Jura, M., 1989, ApJS  70, 329
\bibitem[Kormendy and Richstone(1995)]{bh}Kormendy J., Richstone D., 1995, ARAA, 33, 581
\bibitem[Lavezzi \etal (1999)]{lavezzi} Lavezzi, T. E., Dickey, J. M., Casoli, F., 
       Kaz\'es, I., 1999, AJ, 117, 1995
\bibitem[Lonsdale \etal (1984)]{lonsdale} Lonsdale, C.~J., Persson, S.~E., \& Matthews, K.\ 
         1984, ApJ, 287, 95 
\bibitem[Maloney \& Black (1988)]{maloney} Maloney, P., Black, J.H. 1988, AJ, 325, 389 
\bibitem[Nishiyama \& Nakai (2001)]{nishiyama} Nishiyama, K., Nakai, N., 2001 PASJ, 53, 713
\bibitem[Ochsenbein, Bauer, \& Marcout(2000)]{vizier} 
	Ochsenbein, F., Bauer, P., \& Marcout, J.\ 2000, A\&AS, 143, 23 
\bibitem[O'Sullivan \etal (2001)]{osullivan} O'Sullivan, E., Forbes, D. A., 
Ponman, T. J.,  2001, MNRAS, 328, 461
\bibitem[Paturel \etal(1997)]{leda} Paturel, G., Andernach, H., Bottinelli,
      L., Di Nella, H., Durand, N., Garnier, R., Gouguenheim, L., Lanoix, P.,      
       Martinet,M.C., Petit, C., Rousseau, J., Theureau, G., Vauglin, I., 
	1997, A\&AS, 124, 109 (LEDA - http://leda.univ-lyon1.fr/)
\bibitem[Popescu \etal(2002)]{popescu} Popescu, C.~C., Tuffs, R.~J., 
    V{\" o}lk, H.~J., Pierini, D., \& Madore, B.~F.\ 2002, ApJ, 567, 221 
\bibitem[Roberts \etal(1991)]{roberts} Roberts, M., Hogg, D.E., Bregman, 
	J.N., Forman, W.R., Jones, C., 1991, ApJS, 75, 751
\bibitem[Sage(1993)]{sage93c} Sage, L.\ J.\ 1993, A\&A, 100, 537 
\bibitem[Sage(1993)]{sage93b} Sage, L.\ J.\ 1993, A\&A, 272, 123 
\bibitem[Sanders \& Mirabel (1996)]{sanders} Sanders,D. B., Mirabel, I. F., 1996, 
   ARA\&A, 34, 749 
\bibitem[Solomon \& Sage (1988)]{sage88} Solomon, P. M:, \& Sage, L.\ J.\ 1988, 
    ApJ, 334, 613 
\bibitem[Schweizer \etal(1983)]{rubin3} Schweizer, F., Whitmore, B. C., 
	Rubin, V. C., 1983, AJ, 88, 909
\bibitem[Strong \etal (1988)]{strong} Strong, A.\ W., Bloemen, J. B. G. M., Dame, T. M.  
   et al. 1988, A\&A, 207, 1 
\bibitem[Thronson \& Telesco (1986)]{tronson} Thronson, H. A. Jr., Telesco, C. M.,
   1986, ApJ, 311, 98 
\bibitem[Van Driel \etal (2000)]{vandriel} van Driel, W., Ragaigne, D., 
      Boselli, A., Donas, J., \& Gavazzi, G.\ 2000, A\&AS, 144, 463 
\bibitem[V{\' e}ron-Cetty \& V{\' e}ron(2003)]{veron2003} V{\' e}ron-Cetty, M.-P.~\& 
             V{\' e}ron, P.\ 2003, A\&A, 412, 399 
\bibitem[Vorontsov-Velyaminov(1959)]{vv} Vorontsov-Velyaminov, B.\ A.\ 1959, 
	Atlas and catalog of interacting galaxies (1959), 0 
\bibitem[Welch \& Sage (2003)]{welch} Welch, G. A.; Sage, L. J.,  
2003, ApJ, 584, 260
\bibitem[Whitmore \etal (1990)]{pr} Whitmore, B.C., Lucas, R.A., 
	McElroy, D.B., Steiman-Cameron, T.Y., Sackett, P.D., Olling, R.P., 
	1990, AJ, 100, 1489
\bibitem[Wiklind, Combes \& Henkel(1995)]{wiklind} Wiklind, T., Combes, 
	F.\ \& Henkel, C.\ 1995, A\&A, 297, 643 
\bibitem[Young et al.(1995)]{fcrao} Young, J.\ S.\ Xie Shuding, Tacconi, 
	L. J., et al.\  1995, ApJS, 98, 219 
\bibitem[Zhu et al.(1999)]{zhu} Zhu, M., Seaquist, E.~R., 
	Davoust, E., Frayer, D.~T., \& Bushouse, H.~A.\ 1999, AJ, 118, 145 
\end{thebibliography}
\end{document}